\documentclass[twocolumn]{aastex631}

\newcommand{\dg}{$^\circ$}
\newcommand{\kms}{~km~s$^{-1}$}

\shorttitle{Juno Conjunctions}
\shortauthors{E.E.Davies et al.}
\graphicspath{{./}{figures/}}

\begin{document}

\title{Multi-Spacecraft Observations of the Evolution of Interplanetary Coronal Mass Ejections Between 0.3 and 2.2~AU: Conjunctions with the Juno Spacecraft}

\author[0000-0001-9992-8471]{Emma E. Davies}
\affiliation{Institute for the Study of Earth, Ocean, and Space, University of New Hampshire, Durham, New Hampshire, USA}
\affiliation{Department of Physics, Imperial College London, London, UK}
\correspondingauthor{Emma E. Davies}
\email{emma.davies@unh.edu}

\author[0000-0002-9276-9487]{Réka M. Winslow}
\affiliation{Institute for the Study of Earth, Ocean, and Space, University of New Hampshire, Durham, New Hampshire, USA}

\author[0000-0002-5681-0526]{Camilla Scolini}
\affiliation{Institute for the Study of Earth, Ocean, and Space, University of New Hampshire, Durham, New Hampshire, USA}
\affiliation{CPAESS, University Corporation for Atmospheric Research, Boulder, CO, USA}

\author[0000-0003-2701-0375]{Robert J. Forsyth}
\affiliation{Department of Physics, Imperial College London, London, UK}

\author[0000-0001-6868-4152]{Christian M\"ostl}
\affiliation{Space Research Institute, Austrian Academy of Sciences, Graz, Austria}

\author[0000-0002-1890-6156]{Noé Lugaz}
\affiliation{Institute for the Study of Earth, Ocean, and Space, University of New Hampshire, Durham, New Hampshire, USA}

\author[0000-0003-3752-5700]{Antoinette B. Galvin}
\affiliation{Institute for the Study of Earth, Ocean, and Space, University of New Hampshire, Durham, New Hampshire, USA}

\begin{abstract}

We present a catalogue of 35 interplanetary coronal mass ejections (ICMEs) observed by the Juno spacecraft and at least one other spacecraft during its cruise phase to Jupiter. We identify events observed by MESSENGER, Venus Express, Wind, and STEREO with magnetic features that can be matched unambiguously with those observed by Juno. A multi-spacecraft study of ICME properties between 0.3 and 2.2~AU is conducted: we firstly investigate the global expansion by tracking the variation in magnetic field strength with increasing heliocentric distance of individual ICME events, finding significant variability in magnetic field relationships for individual events in comparison with statistical trends. With the availability of plasma data at 1~AU, the local expansion at 1~AU can be compared with global expansion rates between 1~AU and Juno. Despite following expected trends, the local and global expansion rates are only weakly correlated. Finally, for those events with clearly identifiable magnetic flux ropes, we investigate the orientation of the flux rope axis as they propagate; we find that 64\% of events displayed a decrease in inclination with increasing heliocentric distance, and 40\% of events undergo a significant change in orientation as they propagate towards Juno. The multi-spacecraft catalogue produced in this study provides a valuable link between ICME observations in the inner heliosphere and beyond 1~AU, thereby improving our understanding of ICME evolution. 

\end{abstract}


\section{Introduction} \label{sec:intro}

Coronal mass ejections (CMEs) are large-scale structures of plasma and magnetic field that are expelled from the solar atmosphere. Their heliospheric counterparts, interplanetary coronal mass ejections (ICMEs), undergo many physical processes as they propagate through the heliosphere \citep[e.g.][]{manchester2017physical} which affect their evolution. Understanding this evolution is of great interest in space weather forecasting as they are the main drivers of severe space weather at Earth \citep[e.g.][]{kilpua2017geoeffective}. 

To study ICME properties in-situ, we can use both dedicated solar wind spacecraft e.g. Helios \citep[0.3--1~AU, e.g.][]{cane1997helios, bothmer1998structure}, ACE/Wind, STEREO \citep[at 1~AU, e.g.][]{cane2003interplanetary, richardson2010near, jian2018stereo}, and Ulysses \citep[1--5.4~AU, e.g.][]{liu2005statistical, wang2005characteristics, ebert2009bulk, du2010interplanetary, richardson2014identification}, or planetary mission spacecraft when outside of their planetary environments or during their cruise phase e.g. MESSENGER \citep{winslow2015interplanetary, good2016interplanetary}, Venus Express \citep{good2016interplanetary}, and Juno \citep[1--5.4~AU, e.g.][]{davies2021catalogue} to build ICME catalogues. These catalogues of ICMEs at varying heliocentric distances provide a statistical view of ICME properties and can be used to determine how properties evolve with increasing heliocentric distance as ICMEs propagate \citep[e.g.][]{janvier2019generic}. 

To better understand the magnetic structure and evolution of ICMEs in situ, it is necessary to track individual ICMEs measured over varying scales and separations \citep{lugaz2018}. Previous studies have utilised ICME catalogues to search for conjunctions between different spacecraft that observe the same event; within 1~AU, \citet{good2019self} identified 18 ICME events with clear magnetic flux rope structures and \citet{salman2020radial} identified 47 ICMEs observed by two or more radially aligned spacecraft including MESSENGER, Venus Express, STEREO, or Wind/ACE. Both studies found that the global expansion of the ICMEs with increasing heliocentric distance was consistent with previous statistical trends, but individual events displayed significant variability when compared to average trends.

Such multi-spacecraft catalogues of ICMEs are useful resources and provide the basis for further studies. The multi-spacecraft events identified in \citet{salman2020radial} were utilised by \citet{lugaz2020inconsistencies} to investigate the relationship between the global and local expansion of ICMEs; such measures of local expansion and comparison to global expansion rate for individual ICMEs had not previously been possible, and it was found that the two measures of expansion have little relation to each other. \citet{scolini2021complexity} also used the same multi-spacecraft catalogue to investigate how the magnetic complexity of ICMEs changes as they propagate, and the causes behind such changes; $\sim$65\% of ICMEs were found to change their
complexity between MESSENGER and 1~AU, with interactions between multiple large-scale solar wind structures the main driver of these changes.

More recently, there has been a growing number of missions within 1~AU, including Parker Solar Probe, Solar Orbiter, and BepiColombo. The increasing number of missions in the inner heliosphere provides the opportunity for more alignments between spacecraft and therefore multi-spacecraft observations of ICMEs. Recent case studies of ICMEs observed by Parker Solar Probe and Solar Orbiter in conjunction with spacecraft at 1~AU include \citet{winslow2021first} and \citet{davies2021solo}, respectively. A recent study by \citet{moestl2022multi} presents a list of ICMEs observed in conjunction by Parker Solar Probe, Solar Orbiter, BepiColombo, and spacecraft at 1~AU. As we enter the new Solar Cycle 25 and solar activity increases, we expect to observe many more ICMEs with multi-spacecraft. A live catalogue of such spacecraft conjunctions so far can be found at \url{https://helioforecast.space/lineups}.

Studies of the magnetic structure of ICMEs observed by multi-spacecraft beyond 1~AU remain few: \citet{davies2020radial} investigated a particular ICME event observed by Juno and four spacecraft near the Earth (ACE, Wind, THEMIS B, and THEMIS C) where Juno and the near-Earth spacecraft were separated radially by 0.24~AU and longitudinally by 3.6\dg~and \citet{mulligan1999intercomparison} compared four ICME events observed by the Near Earth Asteroid Rendezvous (NEAR) and Wind spacecraft, with radial separations between 0.18 and 0.63 AU and longitudinal separations between 1.2 and 33.4$^{\circ}$. Multi-spacecraft studies of ICME properties over large heliocentric distances (from 1 to 5.4~AU and beyond) include \citet{nakwacki2011dynamical}, \citet{richardson2014identification}, and \citet{witasse2017icme}. \citet{nakwacki2011dynamical} identified an ICME observed by ACE and later Ulysses at a heliocentric distance of 5.4~AU using magnetic field and plasma data, and \citet{witasse2017icme} identified an ICME at STEREO-A, Mars, comet 67P/Churyumov-Gerasimenko, Saturn, and possibly New Horizons using a variety of instruments including magnetic field, plasma and galactic cosmic ray intensity data. \citet{richardson2014identification} identified 11 possible links between Earth and Ulysses ICME observations based on the direction of propagation and propagation speeds of the ICMEs. This methodology is different to that used in the studies of \citet{davies2020radial}, \citet{mulligan1999intercomparison} and \citet{nakwacki2011dynamical} where ICME observations were matched between spacecraft by identifying features in the magnetic field and available plasma data.  

In this study, we use the \citet{davies2021catalogue} catalogue of ICMEs observed by Juno to compile a database of events also observed by other spacecraft at, or within, 1~AU (MESSENGER, Venus Express, STEREO, or Wind/ACE). Section \ref{sec:identification} presents the methodology used to identify ICME events observed by two or more spacecraft, and Section \ref{sec:msc_database} provides a description of the multi-spacecraft event catalogue and an overview of the separations between observing spacecraft. Section \ref{sec:analysis} presents the analysis of multi-spacecraft events in this study, including the variation of magnetic field strength with increasing heliocentric distance (\ref{subsec:b_variation}), a comparison of local and global measures of expansion (\ref{subsec:expansion}), and an investigation into changes to the orientation of events with magnetic flux ropes as they propagate (\ref{subsec:complexity}). Finally, a summary of our findings is presented in Section \ref{sec:summary}. 

\section{Identification of Multi-Spacecraft Events} \label{sec:identification}

In this study, we make use of the catalogue of ICMEs identified at Juno during its cruise phase in \citet{davies2021catalogue}. The catalogue identified ICME signatures observed in measurements of the magnetic field taken by the Juno Magnetic Field Investigation instrument \citep[MAG;][]{connerney2017juno}. With only magnetic field data available during the Juno cruise phase, ICME candidates were considered only if they met three strict criteria: 1) an enhanced magnetic field magnitude at least twice that of the expected ambient interplanetary magnetic field (IMF) at that distance; 2) a magnetic field profile typical of an ICME, i.e., a shock-like discontinuity, sheath, and region of magnetic ejecta; and 3) a duration on the order of at least one day. It is important to note that with such criteria, the catalogue is biased towards shock-driving ICMEs with strong magnetic field strengths, and therefore is not an exhaustive list of all ICMEs that may be present in the Juno dataset.

During the Juno cruise phase between 2011 August and 2016 July, operational spacecraft within 1~AU included ACE and Wind at L1, STEREO A/B near 1~AU, Venus Express at 0.7~AU and MESSENGER between 0.3 and 0.5~AU. We use the catalogues of ICMEs observed at MESSENGER \citep{winslow2015interplanetary, winslow2017magnetospheric}, Venus Express \citep{good2016interplanetary}, STEREO \citep{jian2018stereo}, Wind \citep{nieves2018understanding}, and ACE \citep{richardson2010near} to help in identifying events observed by multi-spacecraft. With the exception of the ACE catalogue, the ICME events of each catalogue are included in the HELIO4CAST ICME catalog \citep[ICMECATv2.0;][]{moestl2017modeling,moestl2020prediction} with new entries identified by C. M\"ostl. We make use of the HELIO4CAST ICMECATv2.0 catalogue alongside the previous conjunction database of \citet{salman2020radial} to check that no previously identified conjunctions have been missed. 

We use the Juno catalogue as a starting point, and linearly back-propagate (excluding drag) each event using the Propagation Tool (\url{http://propagationtool.cdpp.eu/}) developed at the Institute of Research in Astrophysics and Planetology (IRAP) \citep{rouillard2017propagation} to determine an approximate time-window in which to search for ICME signatures at each spacecraft. With the absence of plasma data, we input an initial estimate range of ICME propagation speed between 350 and 500 \kms, to give an estimate of the earliest time an ICME could likely be observed at a spacecraft, and a more reasonable time, respectively. We also set the maximum angular width of the ICME to be 90\dg~in the ecliptic plane, centred at Juno so as the ICME extends 45\dg~on either side. We choose this width to be consistent with the findings of previous studies, where the likelihood of more than one spacecraft observing the ICME ejecta of an event decreases with increasing longitudinal separation between spacecraft: \citet{good2016interplanetary} found only $\sim$11\% of ICME flux ropes observed at MESSENGER or Venus Express were also observed by another spacecraft whilst longitudinally separated by 45--60\dg. Similarly, \citet{cane1997helios} found only one event of ten observed by both the Helios and IMP8 spacecraft with separations above 40\dg, and \citet{bothmer1998structure} found that for cases with separations $>$60\dg, the magnetic cloud was observed by only one spacecraft. By running the Propagation Tool using an angular width approximately twice this upper limit of longitudinal separation, we aim to capture widely separated events where one or more spacecraft may observe just the flank of the ICME.

By running the back-propagation for each event with these inputs, the Propagation Tool provides an indication of whether each event passed another spacecraft, and the time window in which this may have occurred. We check the magnetic field data of the flagged spacecraft, MESSENGER \citep[MAG;][]{anderson2007magnetometer}, Venus Express \citep[MAG;][]{zhang2007mag}, STEREO A and B \citep[In-situ Measurements of Particles And CME Transients, IMPACT;][]{acuna2008stereo}, ACE \citep[Magnetic Field Experiment, MAG;][]{smith1998ace} or Wind \citep[Magnetic Field Investigation, MFI;][]{lepping1995wind} for signatures of ICMEs during these time periods, complemented by plasma data available at STEREO A and B \citep[In-situ Measurements of Particles And CME Transients, IMPACT;][]{acuna2008stereo}, ACE \citep[Solar Wind Electron Proton Alpha Monitor, SWEPAM;][]{mccomas1998solar} and Wind \citep[Solar Wind Experiment, SWE;][]{ogilvie1995swe}, and further cross-check for entries included in the HELIO4CAST ICME catalogue and the \citet{richardson2010near} catalogue of ICMEs at ACE (\url{http://www.srl.caltech.edu/ACE/ASC/DATA/level3/icmetable2.htm}).

As previously noted, the Juno ICME catalogue is biased towards shock-driving ICMEs. \citet{richardson2010near} previously found that 51\% of ICMEs identified near Earth drove upstream shocks; we therefore recognise that there are likely more ICMEs to be found in the Juno cruise dataset that do not match the criteria previously defined. To search for these ICMEs, we established longer time periods in which other spacecraft were within 45\dg~of longitudinal separation with Juno. We then work forwards from previously catalogued events to establish potential links with signatures in the Juno magnetic data, making sure that average propagation speeds are realistic or in line with speeds observed in situ where plasma data is available. This way, we have attempted to identify in the Juno data all ICMEs identified within 1 AU, whether or not they satisfy the strict criteria described above for which we created our Juno catalogue. We identify an additional 7 ICMEs this way which were not in our initial catalogue as published in \citet{davies2021catalogue}.

Once the potential links between spacecraft within 1~AU and Juno had been established, we attempt to match features in the magnetic field data. With only magnetic field data available during the Juno cruise phase, defining ICME and magnetic ejecta boundaries presented in the Juno ICME catalogue is not straight-forward; only magnetic flux-rope-like structures were considered as magnetic ejecta in the analysis of event properties in \citet{davies2021catalogue}. However, with plasma data available at ACE, Wind and STEREO-A/B, periods of magnetic ejecta can be more confidently defined at these spacecraft; the ICME catalogues of \citet{richardson2010near}, \citet{jian2018stereo}, and \citet{nieves2018understanding} identify regions of low proton temperature to be a reliable indication of magnetic ejecta as well as other plasma signatures \citep[see][]{zurbuchen2006situ}. We make use of the boundaries defined in the HELIO4CAST and \citet{richardson2010near} catalogues to match to magnetic features at Juno and define regions of magnetic ejecta (ME) in this study. We note that these boundaries are therefore different to those defined in \citet{davies2021catalogue}, which have been adjusted to match the boundaries previously defined at other spacecraft where possible.

Further to previous multi-spacecraft studies of ICMEs beyond 1~AU that identify possible associations between ICMEs based only on propagation speed and direction \citep[e.g.][]{richardson2014identification}, we also match features observed in the magnetic data between spacecraft. Identification of ICMEs and their boundaries becomes increasingly difficult with increasing heliocentric distance, and the likelihood of interaction with other ICMEs or solar wind transients also increases \citep[][]{jian2008stream}. The furthest heliocentric distance at which features could be reliably matched between spacecraft occurred whilst Juno was located at $\sim$2.2~AU. This study therefore focuses on multi-spacecraft observations up to 2.2~AU, although there may be other conjunction events with Juno at larger heliocentric distances. 

\begin{figure*}[t!]
\centering
\includegraphics[width = \textwidth]{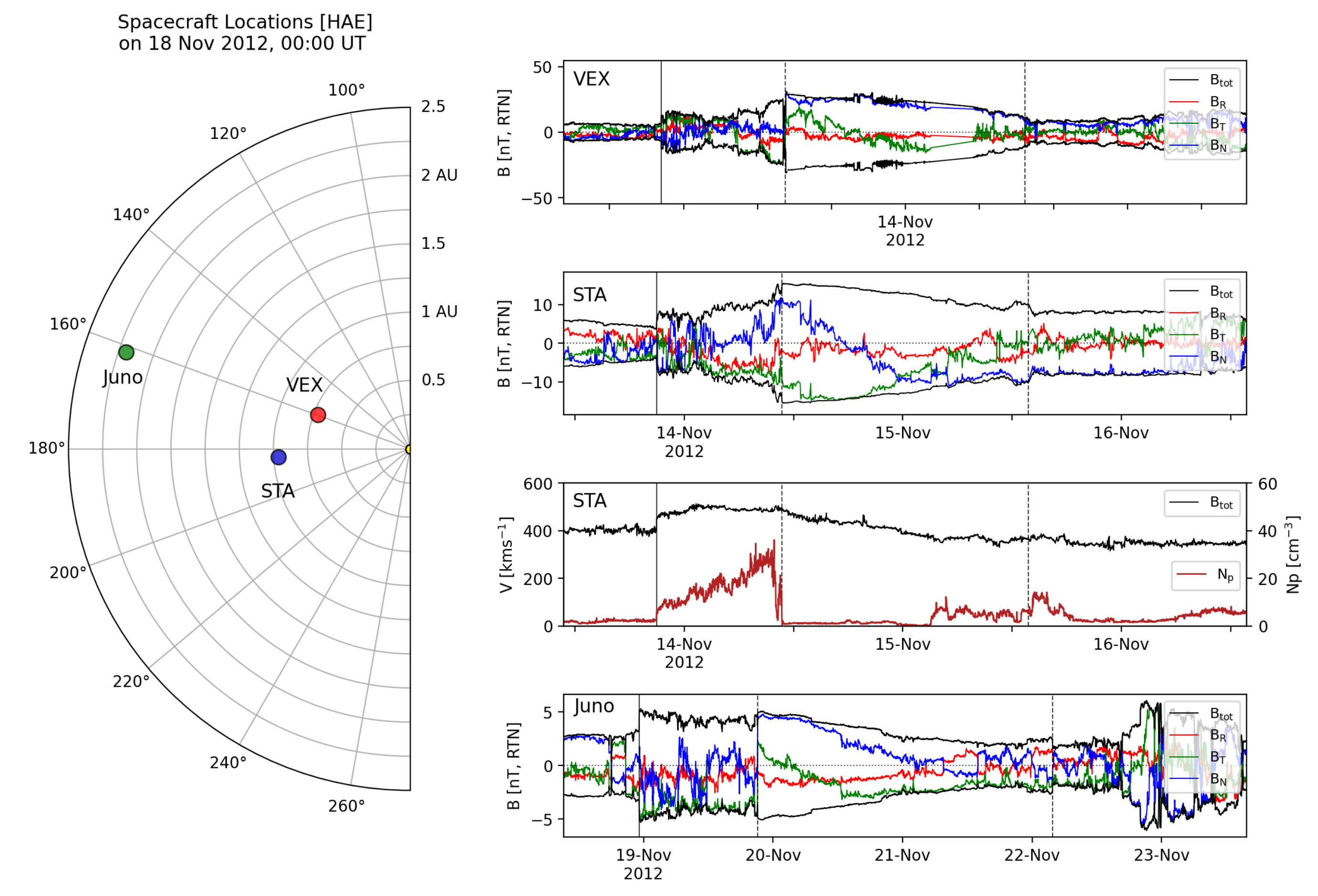}
\caption{Overview of spacecraft positions and in situ data for an example event observed by three spacecraft in November 2012. Left: Spacecraft positions of Venus Express (red), STEREO-A (blue) and Juno (green) on 18 November 2012 in Heliocentric Aries Ecliptic (HAE) coordinates. Right, from top to bottom: magnetic field observations at Venus Express, magnetic field observations at STEREO-A, plasma observations (bulk velocity and proton density) at STEREO-A, and magnetic field observations at Juno. All magnetic field observations are displayed in RTN coordinates (red, green, and blue, respectively). The start of the ICME (the shock-front) is represented by the solid vertical line, and magnetic ejecta (ME) boundaries are delineated by the dashed vertical lines.}
\label{fig:example_event}
\end{figure*}

Figure \ref{fig:example_event} presents an example of one event observed by Venus Express, STEREO-A, and Juno during late November 2012. This event is listed as \#20121118 in both \citet{davies2021catalogue} and this database. The location of each spacecraft are shown on the left-hand side in Heliocentric Aries Ecliptic (HAE) coordinates: Juno is separated radially by 1.47~AU and 1.23~AU with Venus Express and STEREO-A, respectively. Similarly, Juno is separated longitudinally by 8.8\dg~anticlockwise of Venus Express and 18.3\dg~clockwise of STEREO-A during the event. From top to bottom, the right-hand panels display the magnetic field observations (in Radial-Tangential-Normal, RTN, coordinates) at Venus Express and STEREO-A, the plasma observations at STEREO-A (bulk velocity and proton density), and the magnetic field observations at Juno. The magnetic field signatures observed at each spacecraft display a clear flux rope structure with a right-handed rotation of the magnetic field components. The plasma data available at STEREO-A has been useful in defining the boundaries of the magnetic ejecta (especially in the case of the magnetic flux rope, see Section \ref{subsec:complexity}). The ICME observations at Venus Express and STEREO-A are listed in the HELIO4CAST catalog as ICME\_VEX\_SGOOD\_20121113\_01 and ICME\_STEREO\_A\_JIAN\_20121113\_01, respectively. In this case, we have adjusted the boundaries defined in the HELIO4CAST catalog to provide the best match between magnetic features observed across all three spacecraft.

\section{Database of Events Observed by Multi-Spacecraft} \label{sec:msc_database}

\subsection{Description of the Multi-Spacecraft ICME Database}

A total of 35 multi-spacecraft events are identified in our database (published in machine-readable format), 7 of which can be considered triple-alignment events. Table \ref{tab:database} presents a sample of the database. Events are listed chronologically and can be identified by their corresponding identification number (ID\#), which follows the date format `YYYYMMDD'. We use the same ID\#s previously listed in the Juno ICME catalogue for corresponding events, and follow the same system of using the date at which the shock-like discontinuity associated with the event was first observed at Juno for newly identified events in this study. For one event, the ID\# has been given a letter suffix; this corresponds to an event previously noted as a merged interaction region (MIR) in the Juno ICME catalogue that has been resolved to comprise three separate ICME ejectas in this study. Triple alignment events are listed in separate rows under the same ID\#.

For each event in the database, we first list the spacecraft other than Juno at which the ICME was also observed (`Conjunction SC') e.g. MESSENGER (MES), Venus Express (VEX), STEREO-A (STA), STEREO-B (STB) or Wind. For each ICME observed at one of these spacecraft (SC1), three boundaries are defined: ICME start (e.g. a shock or shock-like discontinuity in the magnetic field), ME start, and ME end (columns 3--5). Each datetime is presented in International Organization for Standardization (ISO) format. We list the heliocentric distance (r), latitude, and longitude of the event in HAE coordinates using the ME start time for each event (columns 6--8) and calculate the mean magnetic field strength of the magnetic ejecta (ME B$_{mean}$, column 9). We repeat the same order of columns (10--16) for times and parameters calculated for the same event observed at Juno (referred to as SC2 in the database). 
The final columns (17--19) list velocities calculated for SC1 where plasma data are available (Wind, STEREO-A, and STEREO-B). We list the leading and trailing edge velocities of the magnetic ejecta, calculated by taking a 20 minute average of the bulk velocity following and preceding the defined ME start and end boundaries, respectively. Finally, we perform a linear fit to the bulk velocity over the ME, and take the speed at the midpoint of the ejecta to be the cruise velocity. 

\begin{splitdeluxetable*}{lllllcccBclllcccBcccc}
\tablecaption{Sample of Multi-Spacecraft Events Identified.}
\label{tab:database}
\tablewidth{0pt}
\tablehead{
\colhead{ID\#} & \colhead{Conjunction SC} & \colhead{SC1 ICME Start} & \colhead{SC1 ME Start} & \colhead{SC1 ME End} & \multicolumn{3}{c}{SC1 Position (HAE)} & \colhead{ SC1 ME B$_{mean}$} & \colhead{SC2 ICME Start} & \colhead{SC2 ME Start} & \colhead{SC2 ME End} & \multicolumn{3}{c}{SC2 Position (HAE)} & \colhead{SC2 ME B$_{mean}$} & \colhead{SC1 LE Velocity} & \colhead{SC1 TE Velocity} & \colhead{SC1 Cruise Velocity}\\
\colhead{} & \colhead{} & \colhead{} & \colhead{} & \colhead{} & \colhead{r [AU]} & \colhead{lat $[^{\circ}]$} & \colhead{lon $[^{\circ}]$} & \colhead{[nT]} & \colhead{} & \colhead{} & \colhead{} & \colhead{r [AU]} & \colhead{lat $[^{\circ}]$} & \colhead{lon $[^{\circ}]$} & \colhead{[nT]} & \colhead{[kms$^{-1}$]} & \colhead{[kms$^{-1}$]} & \colhead{[kms$^{-1}$]} 
}
\decimals
\startdata
20110910 & Wind & 2011-09-09 11:46:30 & 2011-09-09 13:53:30 & 2011-09-09 18:19:30 & 1.00 & -0.03 & 346.34 & 18.0 & 2011-09-09 18:01:30 & 2011-09-09 21:18:30 & 2011-09-10 07:13:29 & 1.04 & -0.04 & 352.49 & 16.8 & 436 & 558 & 476\\
20110917 & Wind & 2011-09-17 02:57:00 & 2011-09-17 15:38:30 & 2011-09-18 20:52:30 & 1.00 & -0.01 & 354.17 & 8.2 & 2011-09-17 06:29:30 & 2011-09-17 19:59:30 & 2011-09-18 21:51:31 & 1.07 & -0.05 & 0.79 & 10.4 & 482 & 401 & 444\\
20120518 & STB & 2012-05-12 23:09:00 & 2012-05-13 03:00:00 & 2012-05-14 04:29:00 & 1.00 & 0.19 & 114.96 & 10.6 & 2012-05-17 09:34:30 & 2012-05-18 10:15:30 & 2012-05-20 15:04:31 & 2.14 & -0.02 & 119.04 & 3.7 & 347 & 439 & 415\\
20121118 & VEX & 2012-11-13 10:47:20 & 2012-11-13 17:31:00 & 2012-11-14 06:27:00 & 0.72 & 3.25 & 152.79 & 22.8 & 2012-11-18 23:10:30 & 2012-11-19 21:11:31 & 22/11/2012 03:46 & 2.19 & 0.01 & 161.59 & 3.4 & \nodata & \nodata & \nodata\\	
20121118 & STA & 2012-11-13 20:56:00 & 2012-11-14 10:44:00 & 15/11/2012 13:47 & 0.97 & -0.07 & 179.93 & 14.0 & 2012-11-18 23:10:30 & 2012-11-19 21:11:31 & 22/11/2012 03:46 & 2.19 & 0.01 & 161.59 & 3.4 & 485 & 375 & 426\\
20131029 & MES & 2013-10-29 07:15:28 & 2013-10-29 11:14:46 & 2013-10-29 19:10:55 & 0.33 & -3.29 & 20.63 & 49.1 & 2013-10-29 22:56:31 & 2013-10-31 13:07:31 & 2013-10-31 21:04:30 & 1.10 & 1.95 & 42.10 & 7.4 & \nodata & \nodata & \nodata\\
\enddata
\tablecomments{Each event is identified by an ID number (ID\#). The spacecraft that observed the event in conjunction with Juno is listed (SC1) and the ICME and magnetic ejecta (ME) boundary times in ISO standard. The spacecraft heliocentric distance (r), latitude, and longitude at which SC1 was located are listed in Heliocentric Aries Ecliptic (HAE) coordinates, and the mean ME magnetic field strength. The boundary times, spacecraft position and mean ME magnetic field strength columns are repeated for Juno (SC2). Finally, the average velocities calculated at the leading edge (LE), trailing edge (TE), and the cruise velocity of the ICME are listed. Table \ref{tab:database} is published in its entirety in machine-readable format. A sample is presented here for guidance regarding its form and content.}
\end{splitdeluxetable*}

\subsection{Overview of Multi-Spacecraft Events}

\begin{figure*}[t!]
\centering
\includegraphics[width = \textwidth]{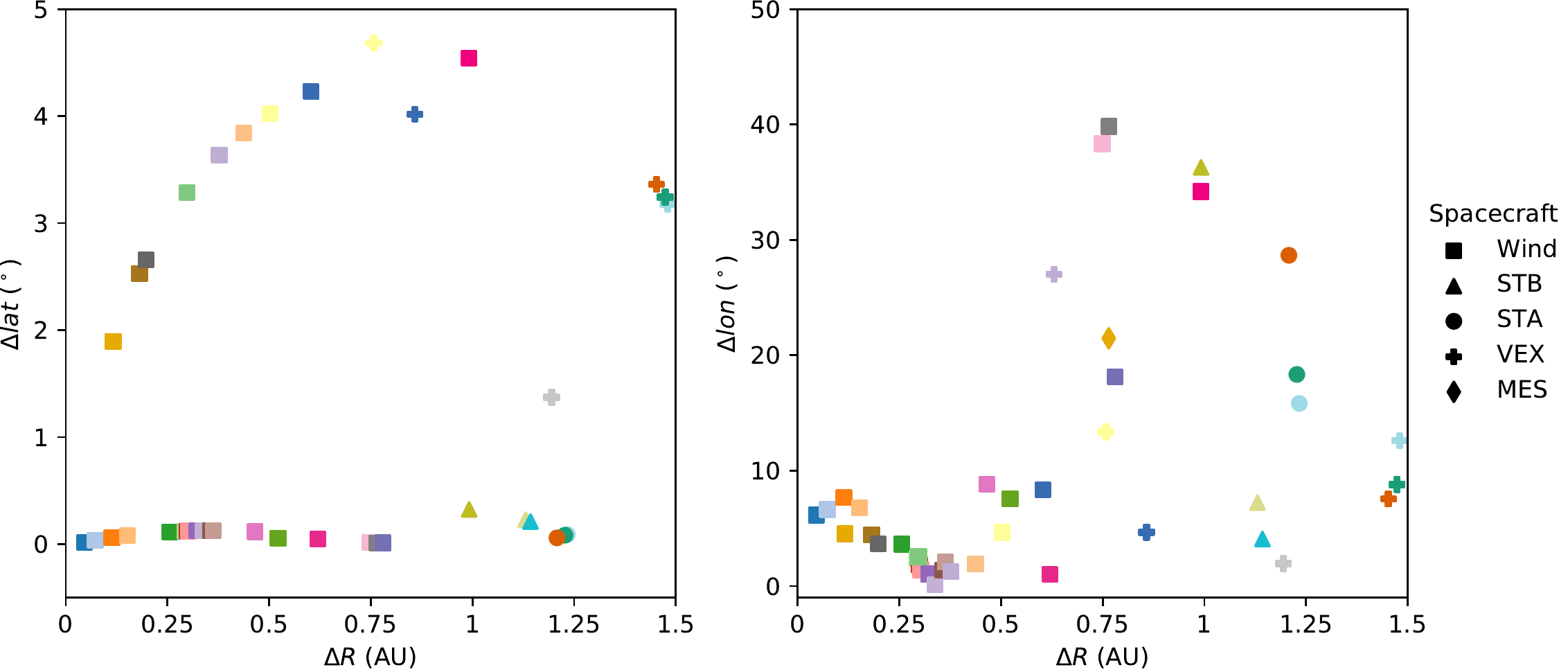}
\caption{Overview of latitude (left) and longitude (right) separations with radial separation between Juno and other spacecraft (SC1) for each event listed in the database in the HAE coordinate system. Each event ID\# corresponds to a unique colour. Marker shapes correspond to the spacecraft other than Juno at which the same event was observed: Wind (square), STEREO-B (triangle), STEREO-A (circle), Venus Express (plus), and MESSENGER (diamond).}
\label{fig:event_separations}
\end{figure*}

Of the 35 multi-spacecraft events, 28 had been previously identified in the Juno ICME catalogue with 7 new events discovered. Of the 7 newly discovered events, we investigate why they were not previously listed in the Juno ICME catalogue: two did not meet the enhancement criterion, three did not meet the general ICME profile criterion, and four did not meet duration requirements of at least one day. 

34 of the 35 events were observed by a spacecraft near 1~AU; 27 at Wind, 4 at STEREO-B, and 3 at STEREO-A. More events were observed by Wind due to the orbit Juno took during its cruise phase which included a gravity assist at Earth, whereas the STEREO spacecraft were already more than 90\dg~from the Sun-Earth line at the time Juno was launched. Of the 27 events observed by Wind, 3 were also observed by Venus Express and one by MESSENGER. All 3 of the events observed by STEREO-A were also observed by Venus Express. The remaining event was observed only by Venus Express before being observed at Juno. 

Figure \ref{fig:event_separations} presents an overview of the latitudinal, longitudinal, and radial separations between each spacecraft and Juno listed in the database (a total of 42 individual conjunctions) in the HAE coordinate system. 29 of the 42 conjunctions took place before the Earth gravity assist performed by Juno in May 2013, where Juno observed events close to the ecliptic plane. With the exception of the planetary mission conjunctions, the latitudinal separations between events are therefore very small in most cases.

The events were observed over a wide range of radial separations, with the smallest separations (from 0.04~AU) occurring shortly after launch and surrounding the gravity assist. The widest separation in heliocentric distance occurs between Venus Express and Juno for event \#20121115, with a radial separation of 1.48~AU. 

Most conjunction events can be considered as close to radial alignment, with 31 of 42 conjunctions (74\%) longitudinally separated by less than 10\dg. In line with previous studies, the number of events observed by both Juno and another spacecraft falls off with increasing longitudinal separation: 12\% of conjunctions are observed at longitudinal separations between 10 and 20\dg, 7\% between 20 and 30\dg, and 10\% between 30 and 40\dg. Despite setting our criterion of 45\dg~when searching for conjunctions, the maximum longitudinal separation found in this study was 39.8\dg. 

The multi-spacecraft event database provides a wealth of interestingly situated events that could be explored further to investigate the global structure of ICME properties, e.g. events with negligible radial and longitudinal separations but wider latitudinal separations, and those with negligible latitudinal and small radial separations that could be used to study the longitudinal differences in ICME structure. Likewise, there are also events with negligible latitudinal and longitudinal separations, with wide heliocentric distance separations that could be explored further to investigate the radial evolution of ICMEs. 

\section{Analysis of Multi-Spacecraft Events} \label{sec:analysis}

\subsection{Variation of Magnetic Field Strength with Heliocentric Distance} \label{subsec:b_variation}

\begin{figure*}[t!]
\centering
\includegraphics[width = \textwidth]{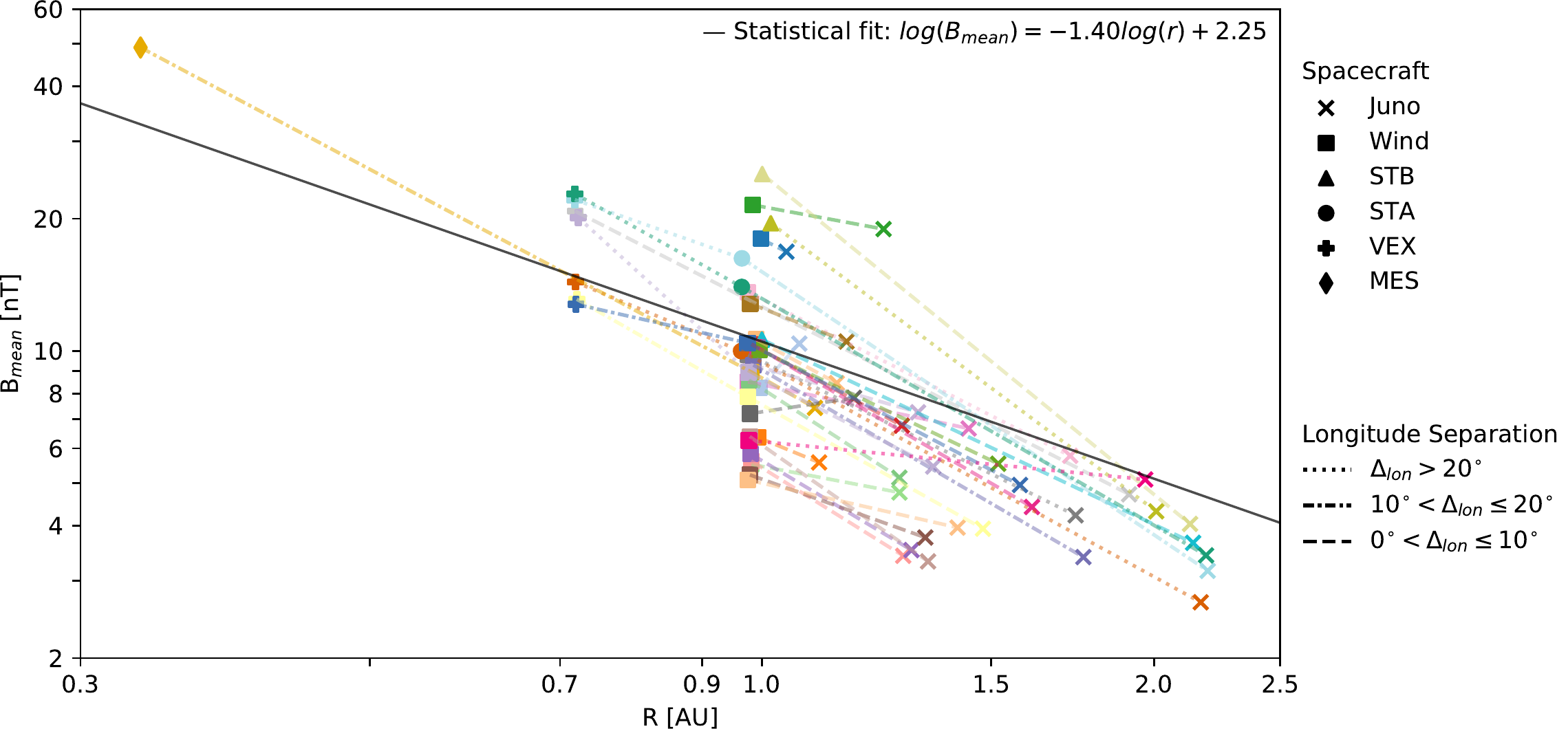}
\caption{Variation with heliocentric distance of the mean magnetic field strengths of each event observed at Juno and at least one other spacecraft. Lines connect measurements of the same ICME, where the type of line used indicates the longitudinal separation between the two spacecraft. Each event is given by a unique colour, the same colour used in Figure \ref{fig:event_separations}. The solid black line represents the statistical best fit through all individual datapoints, the equation for which is given in the top right of the figure.}
\label{fig:B_variation}
\end{figure*}

The variation of ICME magnetic field strength with heliocentric distance provides a measure of their global expansion. Figure \ref{fig:B_variation} presents the mean magnetic field strength of the ICME magnetic ejecta observed at each spacecraft with increasing heliocentric distance on a double-logarithmic scale. Events observed at the different spacecraft display the same symbols and individual colours as displayed in Figure \ref{fig:event_separations}. Lines connect measurements of the same ICME, where the type of line used indicates the longitudinal separation between the two spacecraft: dashed lines indicate longitudinal separations between 0 and 10\dg, dash-dotted lines for separations between 10 and 20\dg, and dotted lines for longitudinal separations greater than 20\dg.

The radial dependence of the magnetic field has been determined by performing separate power law fits for each event \citep[e.g.][]{farrugia2005evolution, richardson2014identification, good2019self, salman2020radial}. Most events display a decrease in magnetic field strength with increasing heliocentric distance between spacecraft. However, there are two exceptions to this trend: events \#20110917 and \#20131111. For event \#20110917, this may be explained by the small radial separation of 0.07 AU between Wind and Juno during this event, but a larger longitudinal separation of 6.6\dg, meaning that significantly different parts of the magnetic flux rope were measured. However, for event \#20131111, the very slight increase in magnetic field strength between Wind and Juno (radial separation of 0.2~AU, longitudinal separation of 3.7\dg) is more likely explained by the interaction with a preceding high speed solar wind transient, where the cruise velocity of the event measured at 1~AU is faster than the speed at the rear of the transient. 

The mean of the fitting parameters for each event gives an average magnetic field strength relationship of $B_{mean} \propto r^{-1.29 \pm 0.95}$, respectively. However, if we remove the events with positive powers mentioned above, we find a relationship of $B_{mean} \propto r^{-1.46 \pm 0.53}$. The uncertainties quoted are the standard deviations, demonstrating the large spread in radial dependencies calculated for individual events. Similarly large variations in radial dependencies have been previously found by \citet{richardson2014identification} for events observed between ACE and Ulysses, and both \citet{good2019self} and \citet{salman2020radial} for events observed by multiple spacecraft between MESSENGER/ Venus Express and spacecraft at 1~AU.

Of the 8 events observed by either MESSENGER and Venus Express, 7 also have conjunctions with spacecraft at 1~AU. We are therefore able to consider the magnetic field relationships within and beyond 1~AU separately. We find for events observed $\le$1 AU, $B_{mean(\le1AU)} \propto r^{-1.51 \pm 0.60}$ and for events observed at distances of $\ge$ 1 AU, $B_{mean(\ge1AU)} \propto r^{-1.25 \pm 1.00}$. The radial dependencies calculated for events observed $\le$1~AU are comparable with the previous studies of \citet{good2019self} and \citet{salman2020radial}: \citet{salman2020radial} found an average maximum magnetic field strength dependency of $B_{mean} \propto r^{-1.75}$ with 50\% of dependencies ranging between $r^{-1.35}$ and $r^{-2.29}$, and \citet{good2019self} found the axial field strength, $B_0$, to have a radial dependency of $B_0 \propto r^{-1.34 \pm 0.71}$. The average $B_{mean(\le1AU)}$ value calculated in this study, lies between the values calculated in those studies. Results of radial field dependencies between ACE and Ulysses (between 1 and 5.4~AU) are most comparable with the $B_{mean(\ge1AU)}$ relationship calculated in this study; \citet{richardson2014identification} found radial dependencies that ranged between $B_{mean} \propto r^{-0.39}$ and $B_{mean} \propto r^{-1.85}$.

Similarly to other studies of ICME expansion, we find that the rate of magnetic ejecta expansion is greater within 1~AU than beyond 1~AU \citep[e.g.][]{leitner2007consequences, lugaz2020inconsistencies, davies2021catalogue}, although not significantly different within uncertainties. If we remove events with positive powers similarly to above, we find $B_{mean(\ge1AU)} \propto r^{-1.45 \pm 0.51}$; this result is similar to $B_{mean(\le1AU)}$. \citet{lugaz2020inconsistencies} theorised that ICME expansion may occur at a similar rate beyond 0.8~AU where it is mostly influenced by the decrease in the solar wind dynamic pressure with increasing heliocentric distance, rather than the initial internal magnetic pressure of the ICME. This theory is perhaps consistent with the similarity of dependencies in this study, where most datapoints (15 of 16) were measured between heliocentric distances of 0.7 and 2.2~AU. 

\citet{richardson2014identification} found that for 11 events observed at Earth and Ulysses, there may be two groups of events: those with strong fields at 1~AU that evolve to weak fields at Ulysses as $r^{-1.5}$ and those with weaker fields at 1 AU that evolve as $r^{-0.7}$. To investigate, we take the highest 25\% of mean magnetic field values ($\langle B_{mean} \rangle$ = 17.6~nT, $\sigma$ = 0.46~nT) and the 25\% lowest mean magnetic field values ($\langle B_{mean} \rangle$ = 5.78~nT, $\sigma$ = 0.48~nT) and compare the corresponding average radial dependencies: -1.63 $\pm$ 0.56, and -1.19 $\pm$ 0.62, respectively. There is a difference in the average radial dependency calculated for each group, although not significantly different within uncertainties nor as distinct as those found by \citet{richardson2014identification}. There is also a large variation of radial dependencies within each group, for example, event \#20111025 has a mean magnetic field strength of 21.5~nT, yet a radial magnetic field dependency of -0.55. This event was previously investigated by \citet{davies2020radial} and found to have been affected by interactions with a following solar wind transient at Wind which was not observed at Juno. It is therefore important to consider that although events with stronger (weaker) magnetic field strengths may expand more rapidly (slowly) in general, the affect of the external solar wind background should be taken into account. 

Considering the ME mean magnetic field measurements at each spacecraft individually, we calculate a statistical relationship of $B_{mean} \propto r^{-1.40 \pm 0.67}$ using a least squares fitting optimisation model (scipy.optimize.least$\_$squares in Python with a loss function of soft$\_$l1), presented by the solid black line in Figure \ref{fig:B_variation}. This relationship is similar to the average magnetic field relationship found for the complete dataset considering only positive powers ($r^{-1.46 \pm 0.53}$). The statistical magnetic field relationship can also be compared to those of ICME studies at Ulysses, which included other ICME datasets at Helios, ACE, and the Pioneer Venus Orbiter to create a dataset covering 0.3--5.4~AU: \citet{richardson2014identification} found $B_{mean} \propto r^{-1.38 \pm 0.03}$, \citet{liu2005statistical} found $B_{mean} \propto r^{-1.40 \pm 0.08}$, and \citet{wang2005characteristics} found $B_{mean} \propto r^{-1.52}$. These powers are very similar to that calculated in this study, despite the difference in heliocentric distance range covered (0.3--2.2~AU in this study), and the lower sample population considered in this study (42 individual spacecraft observations), in comparison to \citet{richardson2014identification} which considers 103 Helios events, $\sim$~300 events at 1~AU from \citet{richardson2010near}, and 279 events at Ulysses.

Finally, we calculate statistical relationships for multi-spacecraft events $\le$1 AU and $\ge$1 AU: $B_{mean(\le1AU)} \propto r^{-1.49 \pm 1.12}$ and $B_{mean(\ge1AU)} \propto r^{-1.29 \pm 0.83}$, respectively. Similarly to above, both relationships agree well with the average relationships calculated. However, in comparison to previous studies that have used similar fitting to derive relationships of mean magnetic field strength with heliocentric distance $\le$1~AU, the expansion calculated is not as fast: \citet{gulisano2010global} found $B_{mean} \propto r^{-1.85 \pm 0.07}$ using Helios 1 and 2 observations, and \citet{winslow2015interplanetary} found $B_{mean} \propto r^{-1.95 \pm 0.19}$ using MESSENGER and ACE observations. This difference is perhaps due to the comparatively small number of datapoints used in this study (16), and their distribution in heliocentric distance with only one datapoint at 0.3~AU. Comparing the $B_{mean(\ge1AU)}$ relationship calculated in this study to previous relationships using Ulysses data between 1 and 5.4~AU, we find the relationships in strong agreement: \citet{ebert2009bulk} found $B_{mean} \propto r^{-1.29 \pm 0.12}$, and similarly, \citet{richardson2014identification} found $B_{mean} \propto r^{-1.21 \pm 0.09}$. The $B_{mean(\ge1AU)}$ relationship is also consistent with that calculated in \citet{davies2021catalogue} of $r^{-1.24 \pm 0.43}$ for ICME magnetic flux ropes between 1 and 5.4~AU, despite the difference in heliocentric distance range (1--2.2~AU). 

\subsection{Comparing Local and Global Measures of Expansion} \label{subsec:expansion}

The dimensionless expansion parameter ($\zeta$) developed by \citet{demoulin2009causes} and \citet{gulisano2010global} provides a measure of the local expansion of an ICME:

\begin{equation} \label{eq:expansion}
    \zeta = \frac{\Delta V}{\Delta t} \frac{r_H}{V_c^2},
\end{equation}

\noindent where $\frac{\Delta V}{\Delta t}$ is the slope of the velocity profile inside the flux rope, $r_H$ is the heliospheric distance at which the measurements are made, and $V_c$ is the speed of the flux rope centre (cruise velocity). 

Typically, values lie in the range $\zeta = 0.81 \pm 0.19$ for non-perturbed magnetic clouds at 1~AU \citep{gulisano2010global}. These authors found that the dimensionless expansion parameter is independent of radial distance and therefore could be used to relate to the global expansion of ICMEs, where the magnetic field strength would be expected to decrease as $r^{-2\zeta}$ and the radial size expected to increase as $r^\zeta$ \citep[see][for discussion]{dumbovic2018analytical}.

We calculate $\zeta$ for each event with plasma data observed at STEREO-A, STEREO-B, and Wind using Equation \ref{eq:expansion}. We find the average $\zeta = 0.60 \pm 0.78$, where the uncertainty quoted is the calculated standard deviation. This uncertainty is particularly large as the values include negative values. However, only including values that indicate positive expansion, we find that the average $\zeta = 0.88 \pm 0.59$. 

\begin{figure}[t!]
\centering
\includegraphics[width = \columnwidth]{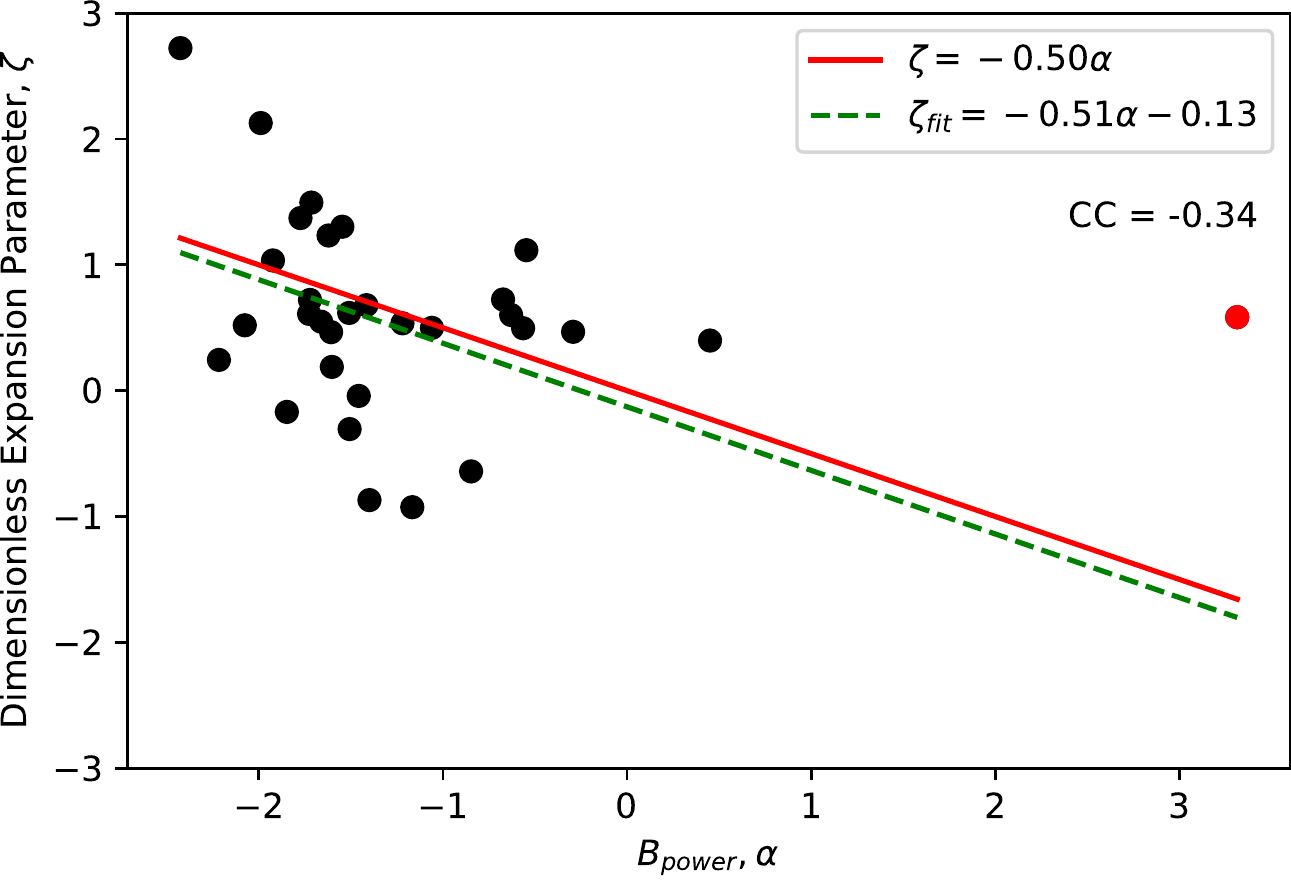}
\caption{Dimensionless expansion parameter at 1~AU, $\zeta$ (local expansion) vs. the radial dependency of the mean magnetic field with heliocentric distance for each event pair, $\alpha$ (global expansion). The dashed green line represents the best linear fit to the data with a gradient of -0.51. This is close to the ideal relationship between the two parameters, with a gradient of -0.5, demonstrated by the solid red line. However, the correlation coefficient (CC) of -0.34 (displayed in the top right hand corner of the figure) shows that the two parameters are weakly correlated.}
\label{fig:dep}
\end{figure}

Figure \ref{fig:dep} presents the variation of the dimensionless expansion parameters calculated at 1~AU for each event, against the radial dependency ($\alpha$) of the mean magnetic field with heliocentric distance for each event pair of the same event. Excluding outliers, we find a best-fit linear relationship between the two parameters of $\zeta_{fit}= -0.51\alpha - 0.13$. This is in close agreement with the expected relationship of $\zeta = -0.5\alpha$. However, the two parameters are weakly correlated, with a correlation coefficient = -0.34. This result is similar to previous studies which have also found a weak correlation between measures of local and global expansion \citep[e.g.][]{lugaz2020inconsistencies}. For values of $\zeta > 1$, the mean radial dependency $\alpha = -1.69 \pm 0.50$, implying the expansion of the magnetic ejecta beyond 1~AU is much faster than the average dependency calculated in Section \ref{subsec:b_variation} ($\alpha = -1.25 \pm 1.00$). For values of $\zeta < 0$, the mean radial dependency $\alpha = -1.39 \pm 0.29$, slightly above average but lower than that of $\zeta > 1$. The widest spread in radial dependencies occurs for values of $\zeta$ between 0 and 1; the mean $\alpha$ in this range is $-1.22 \pm 0.69$. The weak correlation between $\zeta$ and $\alpha$ can therefore be attributed to the large range of expansion rates for the most typical values of $\zeta$ (between 0 and 1), and thus gives little information about the expected decrease in magnetic field strength of events beyond 1~AU. We recognise that the magnetic ejecta of events used in this study differs to the non-perturbed magnetic clouds of \citet{gulisano2010global}, and suggest that the expansion may be dominated by the interaction with the IMF and other transients at 1~AU and beyond.

\subsection{Events with Magnetic Flux Ropes} \label{subsec:complexity}

\begin{figure*}[t!]
\centering
\includegraphics[width = \textwidth]{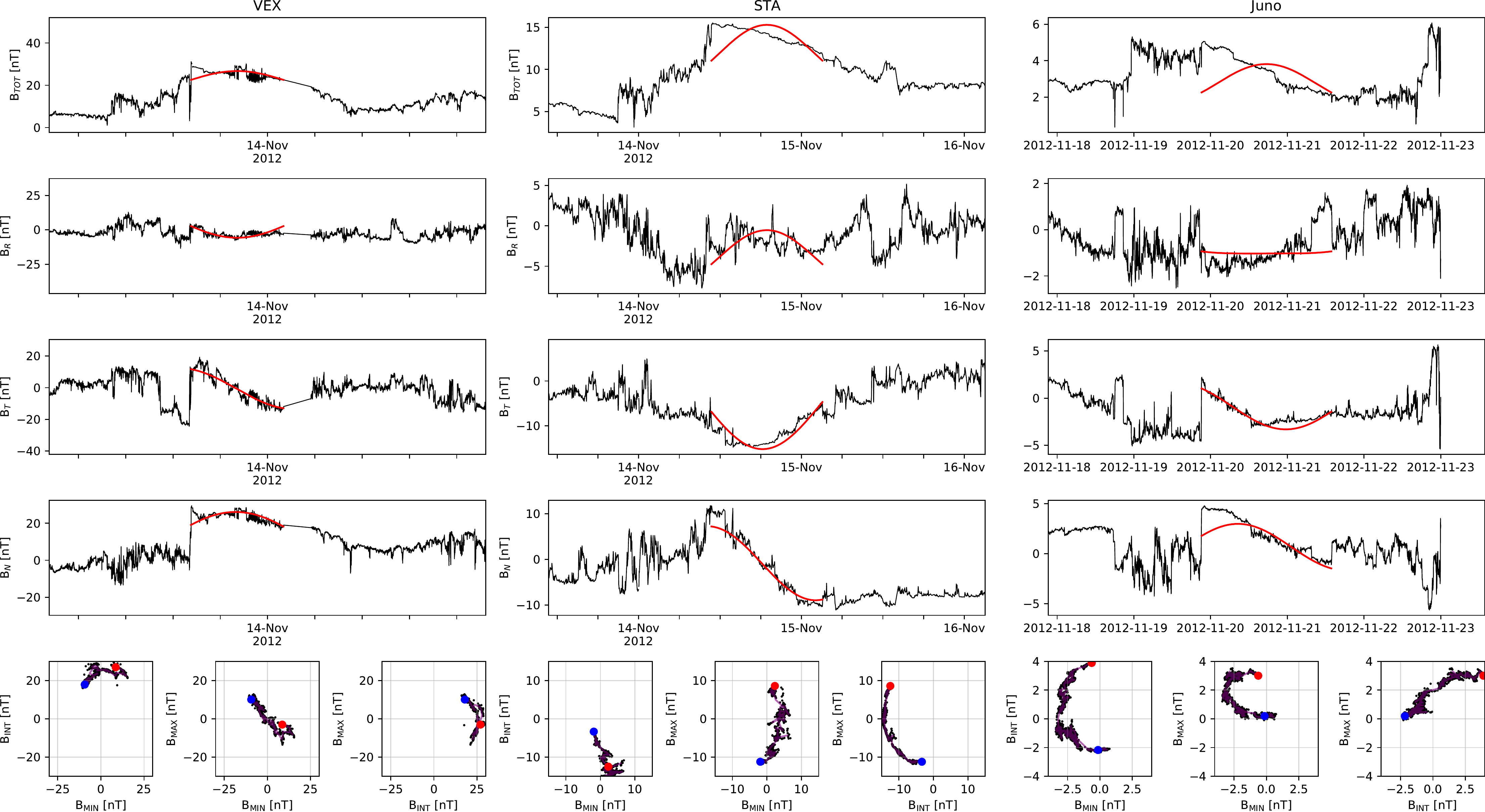}
\caption{The linear force-free (LFF) flux rope fitting results (red) for the same example event presented in Figure \ref{fig:example_event} (event \#20121118) observed at Venus Express (left), STEREO-A (middle) and Juno (right). From top to bottom: the magnetic field magnitude, radial, tangential, and normal magnetic field components (RTN coordinates), and the magnetic hodogram panels in the flux rope coordinate system. The red markers represent the leading edge of the magnetic flux rope, the blue markers represent the trailing edge of the magnetic flux rope, and the purple line is a 10 minute average of the magnetic field data.}
\label{fig:fits_hodograms}
\end{figure*}

As ICMEs propagate, many physical processes can affect their evolution \citep[see][]{manchester2017physical}, including deflections that significantly alter their course \citep{gosling1987deflect} and rotations or changes to their inclination \citep[e.g.][]{nieves2013inner}. Such changes in orientation or reorientations of the magnetic structure can be considered as increases to the magnetic complexity of an ICME \citep[e.g.][]{scolini2021complexity} and have mostly been attributed to their interaction with other solar wind transients \citep{winslow2016longitudinal, winslow2021first, winslow2021effect}. 

Previous studies have found that ICMEs tend to deflect towards the solar equator, particularly during solar minimum when the solar wind is highly structured \citep{plunkett2001solar, cremades2006properties,wang2011statistical}. It has been suggested that ICME magnetic flux ropes may align with the HCS as they propagate \citep{yurchyshyn2008relationship, isavnin2014three}. \citet{isavnin2014three} found that although most of the deflection occurs below 30~R$_S$, a significant amount of deflection and rotation also occurs between 30~R$_S$ and 1~AU. Similarly, \citet{good2019self} observed a small decrease in inclination towards the solar equatorial plane for 77\% of ICME magnetic flux rope pairs observed at MESSENGER, Venus Express or 1~AU with increasing heliocentric distance. In this section, we investigate whether such trends in magnetic flux rope orientation continue beyond 1~AU and whether any significant changes in inclination are present in our database.

To identify events with magnetic flux ropes in our database, we first visually inspect the multi-spacecraft database to identify regions of relatively smooth rotation of the magnetic field components within the defined magnetic ejecta boundaries. For events with spacecraft conjunctions at 1~AU, we also inspect the coincident plasma data for regions of low proton temperature and density, consistent with the definition of a magnetic cloud \citep{burlaga1981magnetic}. 

To confirm the smooth rotation of the magnetic field components, we construct magnetic hodograms of the regions of interest and further refine the selection of data to include those that display a single rotation. Examples of such hodograms are presented in the lower panel of Figure \ref{fig:fits_hodograms}, where the magnetic field data has been transformed into the flux rope frame \citep[e.g.][]{bothmer1998structure} using the eigenvector matrix calculated by performing a minimum variance analysis (MVA). 

The morphological classification scheme of \citet{nieves2018understanding, nieves2019unraveling} can be used to categorise the hodograms into five possible categories: F-, displaying a single rotation lower than 90\dg; Fr, a single rotation between 90\dg--180\dg; F+, a single rotation greater than 180\dg; Cx, includes multiple rotations; and E, an unclear rotation. Each of the F type classifications listed correspond with observations of a spacecraft crossing a single magnetic flux rope structure, where Fr and F- correspond with crossings of a structure comprising a helical magnetic field wrapped around an axis with small and large impact parameters, respectively \citep[][]{nieves2019unraveling}. However, F+ classifications correspond with crossings of structures with significant curvature such as spheromaks \citep{scolini2021swarm} or double magnetic flux ropes \citep{lugaz2013orientation}.

Having refined our magnetic flux rope boundaries to select a single rotation, we further categorise the magnetic flux rope type based on this morphological classification scheme and select only those events comprising flux ropes with a single rotation of less than 180\dg~(F-, Fr) observed at two or more spacecraft to be used in our analysis. We perform a linear force-free (LFF) fit to the magnetic data based on the force-free constant-$\alpha$ flux rope model developed by \citet{burlaga1988magnetic} and later optimised by \citet{lepping1990magnetic}. 

Figure \ref{fig:fits_hodograms} presents the fitting results and hodograms for event \#20121118 (the same event presented in Figure \ref{fig:example_event}) observed at Venus Express (VEX), STEREO-A (STA) and Juno. From top to bottom for each spacecraft, the panels display the magnetic field magnitude, each magnetic field component (RTN), and the magnetic hodogram panels in the flux rope coordinate system where the red marker represents the leading edge value and the blue marker represents the trailing edge. The hodograms give a sense of both the direction and how much rotation of the magnetic flux rope is measured by the spacecraft: from these, we categorise the magnetic flux rope classes as F- at Venus Express, F- at STEREO-A, and Fr at Juno. The results of the LFF fit to the magnetic field components are overlaid in red for each spacecraft. Visual inspection of the fits shows that they fit reasonably well to the data, however, the reduced chi-squared value ($\chi^2_\mathrm{red}$) at Juno is comparatively higher than those calculated at Venus Express and STEREO-A. For observations at STEREO-A and Juno, there is a clear north to south rotation with the axis in the -T direction, which is fit well by the model. For observations at Venus Express, the transverse component of the magnetic field ($B_T$) contributes most to the rotation, with the axis in the +N direction. The largest difference in how the model fits to the data is how it represents the radial component of the magnetic field ($B_R$): at STEREO-A, the $B_R$ component is of opposite sign in comparison to the fits at Venus Express and Juno. This introduces a 180\dg~ambiguity to the magnetic flux rope orientation, which is corrected for by comparing the flux rope axis orientation to the orientation of the magnetic field components at the centre of the flux rope without fitting.

Ten events were identified that met all criteria. Table \ref{tab:complexity} presents the results of the flux rope fitting for each event. Event \#IDs listed correspond to the same event in the catalogue, and the spacecraft at which the events were observed are listed in order of increasing heliocentric distance. The flux rope start and end times are also listed, as previously mentioned, these can differ from the wider magnetic ejecta definition used to define the boundaries given in the database. The radial and longitudinal separations between the spacecraft are also listed. For events with observations at three spacecraft, we first give the separations between the innermost spacecraft and spacecraft at 1~AU, and then those between the spacecraft at 1~AU and Juno. The results from each flux rope fit are presented; these include the magnetic chirality (H), the latitude ($\theta_{0}$) and longitude ($\phi_{0}$) of the flux rope axis (where $\theta_{0}$ is the inclination of the flux rope axis with respect to the R-T plane and  $\phi_{0}$ is the angle of the axis swept out anticlockwise from the Sun-spacecraft line and projected onto the R-T plane), the impact parameter ($y_0$, normalised by the flux rope radius), the magnetic field strength at the flux rope axis ($B_0$), and the reduced chi-squared ($\chi^2_\mathrm{red}$). 

The final columns of Table \ref{tab:complexity} state the change in $\theta_{0}$ and $\phi_{0}$ of the flux rope orientation between spacecraft. We note that although the fitting was conducted in RTN coordinates at each spacecraft, the difference in angular separation between spacecraft is small and thus we assume the R-T plane as equivalent between observations when compared to the large uncertainties of the LFF fitting model. As in \citet{scolini2021complexity}, we define a significant change in flux rope orientation as greater than $\pm 45^\circ$ in $\theta_{0}$ or $\pm 60^\circ$ in $\phi_{0}$. This is in line with the large uncertainties in flux rope orientation produced by the LFF model: for such fits, \citet{lepping2003estimated} estimated errors of $\pm 13^{\circ}$ and $\pm 30^{\circ}$ in $\theta$ and $\phi$, respectively, whereas \citet{alhaddad2013magnetic, alhaddad2018fitting} have previously found that differing fitting models may vary in their estimate of the flux rope axis orientation by $\pm 45^\circ$ in $\theta$ or $\pm 60^\circ$ in $\phi$. Significant changes in orientation are highlighted in bold in Table \ref{tab:complexity}.

The flux rope orientations at each spacecraft were found to have a relatively even split between low and high inclination, where 5/11 events observed at spacecraft at heliocentric distances of 1~AU or less had inclinations of $<$20\dg, and similarly, 5/10 events observed at Juno. Overall, 64\% (7/11) of events displayed a decrease in inclination ($\theta_{0}$) with increasing heliocentric distance, although it is important to note that for 3/7 events, this decrease is $<$5\dg. This result is not quite as significant as the 77\% of events that displayed a decrease in inclination between MESSENGER, Venus Express or 1~AU determined by \citet{good2019self}. The mean change in inclination across all events was found to be a decrease of 3\dg, less than the 8.8~\dg found by \citet{good2019self}. These results indicate that a very small decrease in flux rope inclination may continue to occur beyond 1~AU, however, this result is much smaller than the uncertainties associated with the LFF fitting model.

Four of the ten events (40\%) display at least one significant change in flux rope orientation with increasing heliocentric distance at which it was observed. This is consistent with the 38\% of flux rope events (5/13) found to have significant changes in orientation between MESSENGER and 1~AU by \citet{scolini2021complexity}, however, this result must be taken with care due to the small sample size of events in both studies. We leave the causes behind such changes in orientation as an open question to be investigated by future studies. 

\section{Summary} \label{sec:summary}

We have identified 35 multi-spacecraft events, 28 of which had previously been identified in the Juno ICME catalogue of \citet{davies2021catalogue}. Most of events were observed by Juno and one other spacecraft at 1~AU (Wind, STEREO-A, or STEREO-B), 7 of which were observed by a third spacecraft (MESSENGER or Venus Express). A multi-spacecraft analysis of the in situ magnetic field observations found that:
\begin{itemize}
    \item The mean of the fitting parameters for each event gives an average magnetic field strength relationship of $B_{mean} \propto r^{-1.29 \pm 0.95}$, respectively. However, by removing the events with positive powers (negative expansion), we find a relationship of $B_{mean} \propto r^{-1.46 \pm 0.53}$. The uncertainties quoted are the standard deviations, demonstrating the large spread in radial dependencies calculated for individual events.
    \item We find for events observed $\le$1~AU, $B_{mean(\le1AU)} \propto r^{-1.51 \pm 0.60}$ and for events observed at distances of $\ge$ 1~AU, $B_{mean(\ge1AU)} \propto r^{-1.25 \pm 1.00}$. However, if we remove events with positive powers, we find $B_{mean(\ge1AU)} \propto r^{-1.45 \pm 0.51}$; this result is similar to $B_{mean(\le1AU)}$. Most events (7/8) identified $\le$1~AU were observed at Venus Express (0.7~AU), and therefore we suggest that rates of ICME expansion between 0.7--1~AU and 1--2.2~AU may be similar, consistent with \citet{lugaz2020inconsistencies} where the fastest rates of expansion were theorised to occur within 0.8~AU.
    \item We investigated the dependency of expansion rates beyond 1~AU on the measured magnetic field strength of the ICME at 1~AU by comparing the 25\% highest and lowest values and their corresponding magnetic radial dependencies as they propagate to Juno. We found that ICMEs with stronger magnetic field strengths had an average radial dependency of -1.63 $\pm$ 0.56, and those with weaker magnetic field strengths had an average radial dependency of -1.19 $\pm$ 0.62. Although there is a difference in the average radial dependency calculated for each group, they are not significantly different within uncertainties nor as distinct as those observed by \citet{richardson2014identification} using events between ACE and Ulysses.
    \item Considering the dataset as a whole, we found a statistical relationship of $B_{mean} \propto r^{-1.40 \pm 0.67}$. This relationship is similar to the average magnetic field relationship found for the complete dataset considering only positive powers ($r^{-1.46 \pm 0.53}$). 
    \item Finally, we calculated statistical relationships for multi-spacecraft events $\le$1 AU and $\ge$1 AU: $B_{mean(\le1AU)} \propto r^{-1.49 \pm 1.12}$ and $B_{mean(\ge1AU)} \propto r^{-1.29\pm 0.83}$, respectively. Similarly to above, both relationships agree well with the average relationships calculated. However, in comparison to previous studies that have used similar fitting to derive relationships of mean magnetic field strength with heliocentric distances within 1~AU, the expansion rate calculated is not as rapid. Comparing the $B_{mean(\ge1AU)}$ relationship calculated in this study to previous relationships using Ulysses data between 1 and 5.4~AU, we find the relationships to be in strong agreement.
\end{itemize}

Events observed at spacecraft for which plasma data are available (Wind, STEREO-A, and STEREO-B) are considered in the analysis of the dimensionless expansion parameter, $\zeta$. We investigated the relationship between $\zeta$ and the radial dependency of the mean magnetic field relationships ($\alpha$) for each event, expected to follow $\zeta = -0.5\alpha$ \citep{dumbovic2018analytical}. We found that the two parameters were weakly correlated with a correlation coefficient of -0.34, despite a best fit to the data of $\zeta_{fit} = -0.51\alpha - 0.13$ consistent with the expected relationship.

For events with clearly identifiable magnetic flux ropes, we investigated the orientation of the flux rope axis as they propagated by fitting a LFF model to the magnetic data. 64\% of events were found to display a decrease in flux rope inclination with increasing heliocentric distance, however, for 42\% of those events, the decrease was less than 5\dg. The mean change in inclination across all events was found to be a decrease of only 3\dg, suggesting that a very small change in flux rope inclination may continue to occur beyond 1~AU, however, this result is much smaller than the uncertainties associated with the LFF fitting model. In addition, 40\% of events with magnetic flux ropes underwent a significant change in orientation ($|\Delta\theta|>$ 45\dg or $|\Delta\phi|>$ 60\dg) as they propagated towards Juno, the causes of which are left as an open question for future studies.

The multi-spacecraft catalogue produced in this study provides a valuable link between ICME observations in  the  inner  heliosphere  and  beyond  1  AU, improving our understanding of ICME evolution in situ. Events observed at Juno provide the opportunity to study the radial evolution of ICMEs over large heliocentric distance separations, the most recent opportunity since the Ulysses mission. The investigated distance range in this study is not only interesting for ICME parameters near Earth, but also near Mars.  The multi-spacecraft events identified in this study are intended to be a useful resource for further interesting case studies of ICME evolution over various latitudinal, longitudinal and radial heliocentric distance separations, and to aid in space weather studies at both Earth and other planets.

\begin{acknowledgments}
We have benefited from the availability of the Juno cruise phase data, and thus would like to thank the Juno MAG Principal Investigator, J.E.P. Connerney, and instrument team. We also thank the Planetary Data Science archive (PDS; \url{https://pds.nasa.gov}) for their distribution of Juno data, the NASA Space Physics Data Facility’s Coordinated Data Analysis Web (CDAWeb; \url{https://cdaweb.gsfc.nasa.gov/}) for their distribution of Wind, STEREO, and MESSENGER data, and the ESA Planetary Science Archive (PSA; \url{https://archives.esac.esa.int/psa/}) for the distribution of Venus Express data. This research was supported by funding from the UKRI Science and Technology Facilities Council studentship ST/N504336/1 (E.E.D) and by NASA grants 80NSSC19K0914 (R.M.W. and E.E.D.), 80NSSC20K0431 (E.E.D. and A.G.) and 80NSSC20K0700 (N.L.). C.S. acknowledges the NASA Living With a Star Jack Eddy Postdoctoral Fellowship Program, administered by UCAR's Cooperative Programs for the Advancement of Earth System Science (CPAESS) under award number NNX16AK22G. C.M. thanks the Austrian Science Fund (FWF): P31659-N27, P31521-N27. The HELIO4CAST ICMECAT is available at \url{https://www.helioforecast.space/icmecat} and the multi-spacecraft ICME database of this study is available at \url{https://doi.org/10.6084/m9.figshare.19285385}.
\end{acknowledgments}

\vspace{5mm}
\software{spiceypy \citep{annex2020spiceypy}, scipy \citep{virtanen2020scipy}, scikit-learn \citep{scikit-learn}}


\bibliography{bibliography}{}
\bibliographystyle{aasjournal}


\begin{rotatetable}
 \begin{deluxetable*}{l | l | l l | c c | c c c c c c | c c}
 \tablecaption{LFF fitting results of events with magnetic flux ropes.}
\tabletypesize{\scriptsize}
\centering
 \tablehead{Event \# & SC & FR Start & FR End & $\Delta$r (AU) & $\Delta$lon (\dg) & $H$ & $\theta_0$ ($^\circ$) & $\phi_0$ ($^\circ$) & $y_0$ & $B_0$ (nT)  & $\chi^2_\mathrm{red}$ & $|\Delta \theta_0|$ & $|\Delta \phi_0|$}
 \startdata
 \hline
 20110917 & Wind & 2011-09-17 15:38:30 & 2011-09-18 05:28:30 & & & $-1$ & $37$ & 219 & $0.44$ & 15 & 0.05 & & \\
 20110917 & Juno & 2011-09-17 19:59:30 & 2011-09-18 13:23:31 & 0.07 & 6.62 & $-1$ & $11$ & 309 & $-0.12$ & 14 & 0.05 & 26 & \textbf{90}\\
 \hline
 20111025 & Wind & 2011-10-25 00:28:00 & 2011-10-25 14:21:00 & & & $-1$ & $40$ & 214 & $0.50$ & 28 & 0.04 & & \\
 20111025 & Juno & 2011-10-26 00:40:31 & 2011-10-26 15:27:00 & 0.26 & 3.62 & $-1$ & $25$ & 250 & $0.51$ & 24 & 0.06 & 15 & 36 \\
 \hline
 20120306 & VEX & 2012-03-01 21:26:00 & 2012-03-02 13:21:00 & & & $-1$ & $-11$ & 162 & 0.85 & 35 & 0.03 & & \\
 20120306 & Juno & 2012-03-07 00:32:30 & 2012-03-07 18:44:42 & 1.19 & -1.94 & $-1$ & $-16$ & 175 & 0.85 & 8 & 0.03 & 5 & 13\\
 \hline
 20120512 & STB & 2012-05-08 18:38:00 & 2012-05-09 00:36:00 & & & $1$ & $11$ & 349 & 0.89 & 44 & 0.01 & & \\
 20120512 & Juno & 2012-05-13 15:43:31 & 2012-05-14 07:43:31 & 1.13 & 7.24 & $1$ & $12$ & 348 & 0.96 & 7 & 0.01 & 1 & 1\\
 \hline
 20121115 & VEX & 2012-11-10 22:16:04 & 2012-11-11 00:32:00 & & & $1$ & $23$ & 145 & -0.95 & 41 & 0.01 & & \\
 20121115 & Juno & 2012-11-16 13:37:30 & 2012-11-18 16:35:31 & 1.48 & 7.56 & $1$ & $22$ & 61 & 0.20 & 4 & 0.04 & 1 & \textbf{84}\\
 \hline
 20121118 & VEX & 2012-11-13 17:31:00 & 2012-11-14 01:20:28 & & & $1$ & $8$ & 180 & -0.84 & 43 & 0.01 & & \\
 20121118 & STA & 2012-11-14 10:44:00 & 2012-11-15 03:05:00 & 0.25 & 27.14 & $1$ & $41$ & 175 & 0.66 & 21 & 0.01 & 33 & 5\\
 20121118 & Juno & 2012-11-19 21:11:31 & 2012-11-21 13:46:29 & 1.23 & -18.34 & $1$ & $36$ & 289 & 0.43 & 4 & 0.07 & 5 & \textbf{114}\\
 \hline
 20121129 & VEX & 2012-11-25 05:16:00 & 2012-11-25 11:19:00 & & & $1$ & $7$ & 331 & 0.56 & 20 & 0.03 & & \\
 20121129 & Juno & 2012-11-29 19:03:29 & 2012-12-01 01:09:30 & 1.45 & 7.56 & $1$ & $12$ & 341 & 0.55 & 4 & 0.04 & 5 & 10\\
 \hline
 20130415 & Wind & 2013-04-14 17:02:00 & 2013-04-15 18:08:30 & & & $-1$ & $41$ & 349 & -0.80 & 17 & 0.03 & & \\
 20130415 & Juno & 2013-04-17 01:10:29 & 2013-04-18 13:04:39 & 0.62 & 1.01 & $-1$ & $32$ & 349 & -0.78 & 7 & 0.03 & 9 & 0\\
 \hline
 20130502 & Wind & 2013-04-30 12:43:00 & 2013-05-01 07:12:00 & & & $-1$ & $10$ & 67 & -0.68 & 14 & 0.05 & & \\
 20130502 & Juno & 2013-05-02 10:34:31 & 2013-05-03 06:07:12 & 0.52 & -7.57 & $-1$ & $-6$ & 25 & -0.71 & 8 & 0.03 & 16 & 42\\
 \hline
 20131109 & Wind & 2013-11-09 00:07:30 & 2013-11-09 06:14:30 & & & $-1$ & $-78$ & 69 & -0.51 & 17 & 0.03 & & \\
 20131109 & Juno & 2013-11-09 20:20:30 & 2013-11-09 23:18:30 & 0.18 & -4.45 & $-1$ & $-77$ & 7 & -0.37 & 12 & 0.07 & 1 & \textbf{62} \\
 \hline
 \enddata
 \tablecomments{LFF fitting results for each event with an associated magnetic flux rope. Event ID\#s correspond to those of the database. The start and end times of the flux rope or corresponding boundaries at each spacecraft are listed, and the radial and longitudinal separations between spacecraft observing the same event in order of increasing heliocentric distance are given. The parameters of the LFF fitting are listed: the magnetic chirality ($H$), the latitude ($\theta_0$) and longitude ($\phi_0$) of the flux rope axis, the impact parameter normalized by the flux rope radius ($y_0$), the magnetic field strength at the flux rope axis ($B_0$), and the reduced $\chi^2$ ($\chi^2_\mathrm{red}$). Finally, $|\Delta \theta_0|$ and $|\Delta \phi_0|$ between observing spacecraft in order of increasing heliocentric distance are listed, where values highlighted in bold indicate a significant change in flux rope orientation.}
 \end{deluxetable*}
 \label{tab:complexity}
\end{rotatetable}

\end{document}